\documentclass[prd,twocolumn,showpacs]{revtex4}
\usepackage{epsfig}
\usepackage{graphicx}
\usepackage{amsmath}

\newcommand{\HI}{H$\,${\sc i}}
\newcommand{\LiI}{Li$\,${\sc i}}
\newcommand{\LiII}{Li$\,${\sc ii}}

\begin{document}
\title{Ionizing radiation from hydrogen recombination strongly suppresses the lithium scattering signature in the CMB}

\author{Eric R. Switzer}
\email{switzer@princeton.edu}
\affiliation{Department of Physics, Princeton University, Princeton, New Jersey 08544, USA}

\author{Christopher M. Hirata}
\affiliation{Department of Physics, Princeton University, Princeton, New Jersey 08544, USA}

\date{July 5, 2005}
\begin{abstract}
It has been suggested that secondary CMB anisotropies generated by neutral lithium could open a 
new observational window into the universe around the redshift $z \sim 400$, and permit a 
determination of the primordial lithium abundance.  The effect is due to resonant scattering in the 
allowed \LiI\ doublet ($2s\;^2S_{1/2} - 2p\;^2\!P_{1/2,3/2}$), so its observability depends on the 
formation history of neutral lithium.  Here we show that the ultraviolet photons produced during hydrogen 
recombination are sufficient to keep lithium in the \LiII\ ionization stage in the relevant redshift range and 
suppress the neutral fraction by $\sim 3$ orders of magnitude from previous calculations, making 
the lithium signature unobservable. 
\end{abstract}

\maketitle

\section{Introduction}

There are few observable traces of structure in the universe between the cosmic recombination at redshift $z\approx 1100$ and the
redshifts $z\le 7$ accessible to quasar absorption line measurements \cite{2001AJ....122.2833F}.  Currently our main observational
constraints on this redshift range come from the early integrated Sachs-Wolfe (ISW) effect \cite{1995PhRvD..51.2599H} and the
large-scale Thomson scattering features in the CMB \cite{2003ApJS..148..161K}.  These may be supplemented in the future by
measurements of the \HI\ 21 cm hyperfine line \cite{1997ApJ...475..429M, 2004ApJ...608..622Z, 2004ApJ...613...16F,
2004PhRvL..92u1301L, 2004ApJ...608..611G}.  However, none of these probes can provide detailed information in the redshift
range $200<z<500$: the free electron abundance is too small to produce significant Thomson scattering, the gas is in thermal 
equilibrium with the CMB so there is no 21 cm signal, and the early ISW effect has a blackbody frequency dependence that has no 
redshift information, making the detection of the effect model-dependent.  

The series of papers \cite{2002ApJ...580...29S, 2002ApJ...564...52Z, 2001ApJ...555L...1L} introduces the prospect of using 
anisotropies in the CMB generated by scattering in the resonance lines of neutral lithium to study large scale structure in the 
redshift range $z\approx 200$--$500$.  Because it involves a resonance, the frequency dependence of the \LiI\ scattering feature 
can be mapped into a redshift dependence just as is done for the Lyman-$\alpha$ forest and is proposed for high-redshift 21 cm 
studies.  The frequency dependence deviates strongly from a blackbody, and thus should be distinguishable from the CMB.  This 
idea is also attractive because it provides a potentially clean method of determining the abundance of primordial lithium, before 
it has been destroyed by nuclear processes in stars.  A precision measurement of the lithium abundance from Big Bang Nucleosynthesis (BBN) 
would be an important consistency check of the cosmological model, and would constrain extensions to standard cosmology such as inhomogeneous 
BBN \cite{1985PhRvD..31.3037A, 1986ApJ...309L...1S, 1990ApJ...349..449M}, which tend to increase the $^7$Li abundance.

The primordial abundance of lithium is extremely small, $X_{\mathrm{Li}} \approx 10^{-10}$--$10^{-9}$ \cite{2001PhRvD..63f3512B}, but
the high integrated cross section for excitations of the \LiI\ doublet $2s\;^2S_{1/2} - 2p\;^2\!P_{1/2,3/2}$ suggests the 
possibility of producing a signature within the reach of studies by the {\it Planck}, BLAST, and EDGE experiments -- generally 
experiments with sensitivity around the $300$~$\mu$m band~\cite{2002ApJ...580...29S}. The \LiI\ doublet has an energy of 
$1.847$~eV ($\lambda=6708\,$\AA), so that
the resonant feature would appear to us today at $\lambda = 0.6708(1+z)$~$\mu$m for scattering at redshift $z$.  By comparison,
the lowest single-photon excitation for \LiII\ occurs at 61.3 eV\footnote{The 61.3 eV excitation of ionized lithium is an 
intercombination line, and allowed excitations are $>62$ eV.}, where there are essentially no CMB photons to scatter (and to 
which the pre-reionization universe is opaque due to hydrogen and helium photoionization).  Hence, an observable lithium signature 
on the CMB relies on the formation of neutral lithium after hydrogen recombination.

Previous work on the lithium signature \cite{2002ApJ...580...29S, 2002ApJ...564...52Z, 2001ApJ...555L...1L} emphasizes that with 
a thermal CMB spectrum, lithium recombines around $z=400$ to form a significant neutral fraction.  The neutral lithium gas 
has an optical depth that is high enough to generate a secondary CMB anisotropy, visible today around $268$~$\mu$m.  Important to 
the result is that there are too few ionizing photons in a purely thermal CMB to keep lithium ionized.  A separate line of work 
discusses the non-thermal radiation field generated by hydrogen 
recombination~\cite{1968ApJ...153....1P,1998A&A...336....1B,1993ASPC...51..548R}, which contains ultraviolet (UV) photons from 
both Lyman-$\alpha$ production and $2s \rightarrow 1s$ two-photon decays.  These photons have a maximum energy of $10.1988$~eV, 
nearly double the \LiI\ ionization threshold, $\chi_{\mathrm{Li}} = 5.3917191$~eV \cite{2003IAUJD..17E..13M}.  
Here, we show that this non-thermal ionizing radiation is sufficient to keep lithium almost completely ionized up to late 
times, and to hold the optical depth through the resonance below $10^{-4}$ in the standard cosmology.
  
In this paper, we first calculate the spectral distortion in Sec.~\ref{sec:distortions}, using the three-level approximation for 
hydrogen recombination.  Sec.~\ref{sec:recombphoto} then discusses the photoionization and recombination rates of lithium in the 
presence of the spectral distortion.  In Sec.~\ref{sec:ionscale}, we solve the rate equation for lithium's ionization history, 
and argue that it remains mostly ionized.  Sec.~\ref{sec:signature} outlines the significance this has for a lithium CMB 
signature.  Finally, because this is a non-standard use of the three-level approximation, we justify it subject to some 
possible concerns in Sec.~\ref{sec:subtle}, and sum up the main results in section~\ref{sec:conclusions}.

\section{Atomic hydrogen spectral distortions \label{sec:distortions}}

Neutral lithium can not form if the radiation field contains a sufficient number of photons above the ionization threshold.  This makes
photoionization dependent on the details of high energy tail of the photon distribution ($E \geq \chi_{\mathrm{Li}}$), and the
photoionization cross section above threshold.  The cosmological radiation has nearly a blackbody distribution
\cite{2003ApJ...594L..67F}, except for a small population of non-thermal distortion photons from hydrogen recombination in the high
energy tail of the distribution \cite{1968ApJ...153....1P,1998A&A...336....1B,1993ASPC...51..548R}.

The most important sources of non-thermal radiation during hydrogen recombination are the \HI\ two-photon transition ($2s 
\rightarrow 1s$) and the \HI\ Lyman-$\alpha$ transition ($2p \rightarrow 1s$).  Very few distortion photons
are produced in the higher-order Lyman lines because such photons are efficiently destroyed by absorption followed by decay to an
excited state of \HI.  Also, the Balmer, Paschen, etc. lines of \HI, as well as nearly all of the photons produced by direct
recombination to \HI\ excited states have energies below 5.39 eV, and so cannot ionize \LiI.  The populations of the $2s$ and 
$2p$ states are expected to vary out of equilibrium at late times.  The bearing this has on the formation history of 
neutral lithium is discussed in Sec.~\ref{sec:subtle}.

We will see that the non-thermal photons from $n=2 \rightarrow 1$ decays of \HI\ have an abundance of $10^{-9}$ relative to the 
the peak of the thermal distribution, but exceed the CMB photon population above the ionization threshold of lithium 
\cite{1998A&A...336....1B}, on the high-energy tail of the CMB.  The ionizing radiation from these decays must be included in a 
calculation of lithium's ionization history.

A natural quantity to track is the number of photons per hydrogen, per log-energy.  Written in 
terms of the phase space density ${\cal N}$ and specific intensity $J_\nu$, 
\begin{equation}
r \equiv \biggl [ \frac{ \mathrm{photons} }{ \mathrm{ hydrogen } \cdot \ln{(E)} }
\biggr ]  = \frac{ 8 \pi E^3 }{ h^3 c^3 n_H } {\cal N} = \frac{ 4 \pi }{ h c n_H } J_\nu,
\label{eqn:rdistortion}
\end{equation}
where $n_H$ is the hydrogen number density and $h$ and $c$ are Planck's constant and the speed of light, respectively.
This choice also has the advantage that it is conserved along a trajectory, unlike $J_\nu$.  The 
photon number densities and baryon number densities both dilute by $(1+z)^3$.  With these 
units, the high-energy Wien tail of the blackbody distribution scales as, 
\begin{equation}
r(E,z) \approx \frac{ 8 \pi E^3 }{ h^3 c^3 n_H } e^{-E/(k T_r)}.
\label{eqn:CMBphotons}
\end{equation}
(After $z \approx 200$, matter and radiation are decoupled, so their thermal histories depart 
and must be tracked separately as a radiation temperature $T_r$, and matter temperature $T_m$.) 
From the integral over the asymptotic distribution, there are $\approx 80$ photons capable of ionizing 
lithium, per lithium atom at $z=500$, and only $\approx 1\times 10^{-11}$ at $z=300$ from the thermal 
distribution.  Purely thermal CMB radiation is not enough to keep lithium ionized after 
$z=400$, but the distortion extends to much higher energies.  

We find the distortion by calculating the rates of two-photon and Lyman-$\alpha$ 
decays, and transport the spectrum of photons that is generated from reactions of these rates 
and the known emission profiles.  The method works {\it post-facto} because it takes the 
ionization history of hydrogen and its derivatives, and then calculates the radiation field.  
Ideally, the evolution of the radiation field would be included in the recombination calculation, 
self-consistently, rather than computed later.  Results from a recombination 
calculation with a self-consistent radiation field are expected to be different at the percent 
level because of feedback and other effects~\cite{2000ApJS..128..407S}.  Here, we want to determine 
whether lithium remains ionized, and do not need anything so nuanced.  

Assuming that only the two $n=2$ decays of hydrogen to its ground state contribute to the 
distortion, all the rate information can be extracted from the total hydrogen recombination 
rate in the three-level approximation~\cite{1968ApJ...153....1P, 2000ApJS..128..407S},
\begin{eqnarray}
\frac{ dx_e }{ dz } &=& \frac{ x_e^2 n_H \alpha_H - \beta_H (1-x_e) e^{-E_{Ly\alpha}/(k T_m)} }{
(1+z) H(z) } \nonumber \\
&& \times \frac{\Lambda_{2\gamma} + 1/ (K n_H (1-x_e)) }{ \Lambda_{2\gamma} + \beta_H + 1/(K 
n_H (1-x_e))},
\label{eqn:hydrogenthreelevel}
\end{eqnarray}
where $x_e$ is the fraction of electrons relative to hydrogen, $\alpha_H$ is the case B recombination rate of 
hydrogen, $\beta_H$ is the effective case B photoionization rate,
\begin{equation}
\beta_H = \alpha_H e^{-\chi_{2s}/(k T)} \frac{ ( { 2 \pi m_e k T } )^{3/2} }{ h^3 },
\label{eqn:hydrogenphoto} 
\end{equation}
$\Lambda_{2\gamma}$ is the $2s \rightarrow 1s$ two-photon decay rate, $\chi_{2s}$ the ionization energy from the $2s$ state, and 
$K=\lambda_{Ly\alpha}^3 /[8 \pi H(z)]$ is the Peebles inverse resonance escape rate from the Lyman-$\alpha$ line.  The form of the 
three-level rate equation is suggestive: the term on the second line is the fraction of excited hydrogen atoms that make the 
transition $2p\rightarrow 1s$ or $2s \rightarrow 1s$, relative to the atoms that are photoionized.  The term in the numerator 
displays a two-photon piece and a Lyman-$\alpha$ piece, while the denominator also has a contribution from photoionization.  This 
makes it easy to decouple the relative contributions of each process to the overall hydrogen recombination rate: call the 
fractional rate of change in $x_e$ during hydrogen recombination due to two-photon and Lyman-$\alpha$ transitions 
$(dx_e/dz)|_{2\gamma}$ and $(dx_e/dz)|_{Ly\alpha}$, respectively.
 
The solutions to the three-level rate 
equation and evolution of the matter temperature are evaluated by {\verb recfast }, developed 
in Seager et al.~\cite{1999ApJ...523L...1S,2000ApJS..128..407S}.  We have modified it to return the 
fractional rates, $(dx_e/dz)|_{2\gamma}$ and $(dx_e/dz)|_{Ly\alpha}$ as a function of $z$.  Throughout, 
the calculations are based on the cosmology $\Omega_m = 
0.27$, $\Omega_b = 0.046$, $H_0 = 70$~km~$\mathrm{s}^{-1}$~$\mathrm{Mpc}^{-1}$, $T_{\mathrm{CMB}} = 2.728$~K, and $Y_{He} = 0.24$.  

\begin{figure}
\epsfxsize=3.3in
\begin{center}
\epsffile{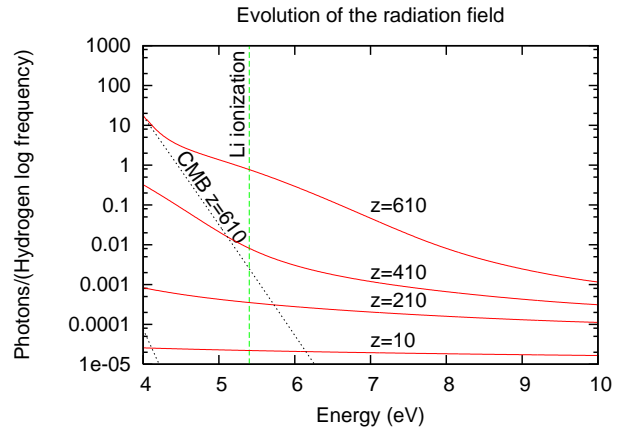}
\end{center}
\caption{The distortion field generated by $n=2$ decays of hydrogen during recombination dwarfs 
the high-energy tail of the CMB (shown in black dotted line).  The ionization energy of lithium is 
shown to emphasize that even at late times, there is a large number of distortion photons 
above the ionization energy.}  
\label{figs:radevo} \end{figure}

To understand how photons generated during hydrogen recombination influence the ionization
history of lithium, they must be transported to later redshifts.  Since Thomson scattering does not
significantly change the photon energy at the redshifts of interest\footnote{When a photon scatters off an electron, it loses
a fraction of order $\sim E/m_ec^2$ of its energy due to recoil, so the condition for recoil from repeated
scatterings to be negligible is $\tau\ll m_ec^2/E$, where $\tau$ is the optical depth.  In this case, the distortion photons have 
energy $E\sim 10\,$eV and $m_ec^2=511\,$keV, so $m_ec^2/E\sim 5\times 10^4$.  By comparison, distortion photons are first 
produced around redshift $z\sim 1500$, and the optical depth to this redshift is $\sim 30$.  Recoil can be neglected.  Likewise, 
the change in energy due to the Doppler effect over many scatterings is $\Delta E/E \sim \sqrt{\tau} \sqrt{(kT)/m_e 
c^2}$, yet $\tau \ll m_e c^2 /(kT)$, so Doppler effects can be neglected.}, transport amounts to redshifting 
the emitted distributions.  The set of distortion photons around some energy $E$ is the sum over all photons which were
produced at an earlier time ($z_{\mathrm{em}}$) with energy $E_{\mathrm{em}}$ which could 
have redshifted to the energy $E$ by redshift $z$.  This defines an integral along the line of sight of 
the contribution of a particular process to the radiation field, 
\begin{equation}
r_{\mathrm{process}}(E) = \int_z^{z_{\mathrm{end}}} dz_{\mathrm{em}} E_{\mathrm{em}} 
\phi(E_{\mathrm{em}}) \frac{ dx_e }{ dz} (z_{\mathrm{em}}) \biggr |_{\mathrm{process}},
\label{eqn:transportlineofsight} 
\end{equation}
where $\phi(E_{\mathrm{em}})$ is the emission profile of the process, evaluated at the 
energy of the radiation at redshift $z_{\mathrm{em}}$ that will have redshifted down to 
an energy $E$ by redshift $z$,  
$E(z)_{\mathrm{em}} = E (1+z_{\mathrm{em}})/(1+z)$.
The combination of the profile and frequency suggests a new quantity:
\begin{equation} 
\psi(E_{\mathrm{em}}) = E_{\mathrm{em}} \phi(E_{\mathrm{em}}) = \biggl [ \frac{ \mathrm{photons} 
}{ \mathrm{decay} \cdot \ln(E_{\mathrm{em}}) } \biggr ].
\label{eqn:profile}
\end{equation}
The integral Eq.~(\ref{eqn:transportlineofsight}) along a line of sight is equivalent to an 
integral over energy under the change of variables:
\begin{equation}
E_{\mathrm{em}} = E \frac{(1+z_{\mathrm{em}}) }{ (1+z)} \Rightarrow dz_{\mathrm{em}} = 
(1+z_{\mathrm{em}}) \frac{ dE_{\mathrm{em}} }{ E_{\mathrm{em}}} .
\label{eqn:lineofsightvariables}
\end{equation}
The integral along the line of sight adds contributions from increasingly higher redshifts, with higher emission energies 
$E_{\mathrm{em}}$.  Because the emission profile is the same as time goes on, this is equivalent to adding up different 
contributions over the profile evaluated at progressively higher energies.  The change of variables amounts to converting this 
indirect way of summing over redshifts into a direct, convenient sum of the profile and rates over energy,  
\begin{equation}
r_{2 \gamma}(E) = \int_E^{E_{Ly \alpha}} (1+z_{\mathrm{em}}) \frac{ dE' }{ E' } 
\psi_{2\gamma} (E') \frac{ dx_e }{ dz } (z_{\mathrm{em}}(E')) \biggr |_{2 \gamma }. 
\label{eqn:twophotonintegral} \end{equation}
Here, $z_{\mathrm{em}}$ is the redshift at which light was emitted with energy 
$E_{\mathrm{em}}$ that will redshift to energy $E$ at some later redshift $z$,
\begin{equation}
z_{\mathrm{em}}(E) = \frac{E_{\mathrm{em}} }{ E} (1+z) -1.
\label{eqn:escapez}
\end{equation}
We use the polynomial fit to the Spitzer \& Greenstein~\cite{1951ApJ...114..407S} two-photon
profile provided in~\cite{2003adu..book.....D}, and assume a delta function for the 
Lyman-$\alpha$ profile.  Results from the calculation are shown in Fig.~\ref{figs:radevo}.

\section{Lithium recombination and photoionization \label{sec:recombphoto}}

The photoionization rate of \LiI\ has a thermal contribution from the CMB and an non-thermal 
contribution from the integral of the cross section over the distortion field, 
$\beta_{\mathrm{Li}} = \beta_{\mathrm{CMB}} + \beta_{\mathrm{dist}}$.  The thermal 
photoionization rate is related by the Milne relations to the recombination rate 
$\alpha_{\mathrm{Li}}$,
\begin{equation}
\beta_{\mathrm{CMB}} = \frac{ g_e g_{\mathrm{LiII}} }{ g_{\mathrm{LiI}} }\alpha_{\mathrm{Li}} (T_r) \exp \biggl \{ 
- \frac{ \chi_{LiI} }{ k T_r } \biggr \} \frac{ ( 2 \pi m_e k T_r )^{3/2} }{ h^3 }.
\label{eqn:lithermalphotoionization}
\end{equation}
The Milne relations assume Boltzmann-distributed excited states at temperature $T_r$.  Yet, the distortion
photons may increase the population of excited states, increasing the ionization rate from CMB photons.  If this effect is 
important, it can only increase $\beta_{\mathrm{CMB}}$, and thus reduce the neutral lithium abundance.  Ionization by a distortion 
photon occurs predominantly from the ground state.  For the case A \LiII$\rightarrow$\LiI\ recombination 
rate, $\alpha_{\mathrm{Li}}$, we use a fit from Verner and Ferland~\cite{1996ApJS..103..467V},
\begin{equation}
\alpha_{\mathrm{Li}}(T_m) = a \biggl [ \sqrt{\frac{T_m }{ T_0}} \biggl (1 + \sqrt{\frac{T_m}{T_0}}
\biggr )^{1-b} \biggl( 1 + \sqrt{\frac{T_m }{ T_1}} \biggr )^{1+b} \biggr ]^{-1},
\label{eqn:liradiativerecombination}
\end{equation}
where the coefficients are, $a=1.036\times 10^{-11}\,$cm$^3\,$s$^{-1}$, $b=0.3880$, $T_0 =
1.077\times 10^{2}$~K, and $T_1 = 1.177 \times 10^{7}$~K.  The fit is accurate to better than
4\% (with maximum error less than 6\%)~\cite{1996ApJS..103..467V} between $3$~K and $10^9$~K.

\begin{figure}
\epsfxsize=3.3in  
\begin{center}
\epsffile{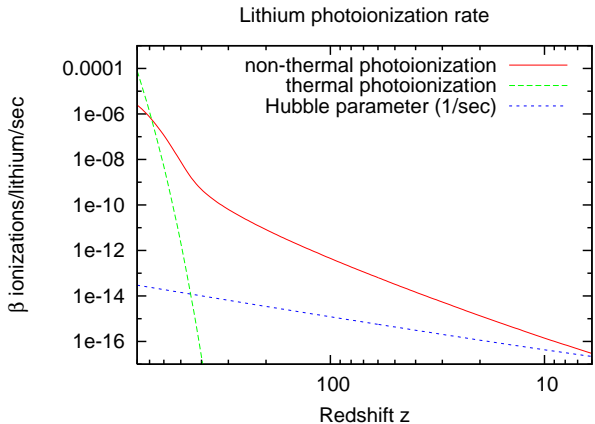}
\end{center}
\caption{The photoionization rate of lithium as a function of redshift, showing the contribution 
of non-thermal (distortion) photons, beginning to dominate shortly after hydrogen 
recombination.  The non-thermal rate is fast compared to the inverse Hubble time, allowing a steady-state solution.} 
\label{figs:photoionization} \end{figure}

The non-thermal contribution to lithium photoionization from photons emitted during hydrogen
recombination is the integral over the cross section,
\begin{equation}
\beta_{\mathrm{dist}} = \int_{\chi_{LiI}}^{E_{Ly\alpha}} \frac{dE'}{E' } \biggl (r_{2 
\gamma}(E') + r_{Ly \alpha} (E') \biggr ) \sigma_{\mathrm{Li}}(E') c.
\end{equation}
We use a fit to the photoionization cross section for \LiI\ from \cite{1996ApJ...465..487V} that has been smoothed over
resonances.  (There are no resonances in our energy range, so this is not a concern.)  The fit is based on low-energy cross 
sections from the Opacity Project~\cite{1993BICDS..42...39C} and high-energy cross sections from a Hartree-Dirac-Slater method 
(though the high-energy cross section is unimportant here because there are no photons above $E_{Ly\alpha} = 10.2$~eV): 
\begin{equation}
\sigma(y=E/E_0) = \sigma_0 (y - 1)^2 y^{0.5P-5.5} ( 1 + \sqrt{y/y_a})^{-P},
\label{eqn:lithiumphotoionizationcrosssection}
\end{equation}
where $\sigma_0 = 1.87723 \times 10^{-6}$~$\mathrm{cm}^3$ $\mathrm{s}^{-1}/c$, $P = 4.895$, and $y_a = 15.01$.  The
results for the non-thermal and thermal contributions to lithium photoionization as a function of
redshift are shown in Fig.~\ref{figs:photoionization}.

\section{Ionization history and scaling \label{sec:ionscale}}

Lithium chemistry and the interaction between lithium ions, lithium hydride, etc. can be very 
complicated \cite{1998A&A...335..403G}, but we will follow Stancil et al.~\cite{2002ApJ...580...29S} in 
tracking only photoionization and radiative recombination, the two most dominant reactions,
\begin{equation}
\mathrm{Li}^+ + \mathrm{e}^- \leftrightarrow \mathrm{Li} + \gamma.
\label{eqn:reactions}
\end{equation}
The next leading order reaction is radiative charge transfer~\cite{1996ApJ...472..102S, 
2002ApJ...580...29S},
\begin{equation}
\mathrm{Li} + \mathrm{H}^+ \rightarrow \mathrm{Li}^+ + \mathrm{H} + 
\gamma,
\label{eqn:chargetransfer}
\end{equation}
which is two orders of magnitude slower than Eq.~(\ref{eqn:reactions}) if distortion photons are neglected.  Photoionization from 
distortions will only speed up Eq.~(\ref{eqn:reactions}) relative to
Eq.~(\ref{eqn:chargetransfer}), making charge transfer less important.
 
\begin{figure}
\epsfxsize=3.3in  
\begin{center}
\epsffile{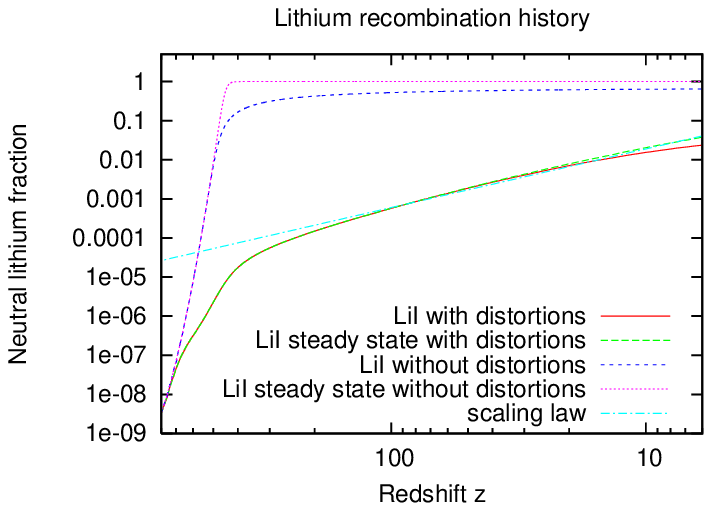}
\end{center}
\caption{Lithium recombination history in the presence of an ionizing flux from the $n=2$ 
hydrogen recombinations to the ground state.  Lithium recombines much more slowly than for a 
purely thermal CMB, as $(1+z)^{-3/2}$, and remains largely ionized up until late times.  This 
reduces the optical depth, and significantly reduces the imprint left on the CMB.} 
\label{figs:lirec}
\end{figure}

The rate equation describing the evolution of the neutral lithium fraction $f_{\mathrm{Li}}$  based on the recombination 
$\alpha_{\mathrm{Li}}$ rate and photoionization rate $\beta_{\mathrm{Li}}$ is 
\begin{equation}
\frac{df_{\mathrm{Li}} }{ dz} = -(1+z)^{-1} H(z)^{-1}[\alpha_{\mathrm{Li}} n_e 
(1-f_{\mathrm{Li}}) - \beta_{\mathrm{Li}} f_{\mathrm{Li}}]. \label{eqn:rateequation}
\end{equation}
The solution of the rate equation Eq.~(\ref{eqn:rateequation}) is shown in 
Fig.~\ref{figs:lirec}.

Lithium recombination according to Eq.~(\ref{eqn:rateequation}) goes through four major stages:
\newcounter{LiCt}
\begin{list}{\arabic{LiCt}. }{\usecounter{LiCt}}
\item A Saha stage, where the lithium ionization states are in thermodynamic 
equilibrium;
\item A primary steady-state stage, in which the spectral distortion from the bulk of hydrogen recombination at $z\sim 1300$ keeps 
lithium ionized;
\item A residual steady-state stage, in which the spectral distortion from $z\sim 1300$ has redshifted to below the \LiI\ ionization 
threshold, but lithium is kept ionized by UV radiation from residual hydrogen recombinations; and
\item A freeze-out stage, in which the time scale for all relevant reactions becomes larger than the Hubble time, and 
$f_{\mathrm{Li}}$ approaches a constant.  It is during this stage that the steady-state approximation, 
Eq.~(\ref{eqn:steadystate}), breaks down. 
\end{list}
In the real universe, the first stars eventually turn on, increasing the UV background--probably ionizing essentially all of the 
intergalactic lithium.  This occurs too late (i.e. too low $z$) to be of interest for the lithium feature in the CMB.

We now discuss each of the four stages in more detail.

\subsection{Saha stage ($z>700$)}

Initially, \LiI\ and \LiII\ co-exist in a Saha
equilibrium, where many thermal ionizing photons are present and the distortion is negligible.  Note
that the neutral fraction during this phase is extremely small, $f_{\mathrm{Li}}<10^{-7}$.  The Saha
equation in this case is
\begin{equation}
f_{\mathrm{Li}} = n_e \left( \frac{h^2}{2\pi m_e k T_r} \right)^{3/2} e^{\chi_{LiI}/(k T_r)}.
\end{equation}
This equation is valid until $z\sim 700$.

\subsection{Primary steady-state stage ($400<z<700$)}

Around $z\sim 700$, enough CMB photons redshift below the \LiI\ ionization threshold so that
a small number of distortion photons dominate the \LiI\ photoionization rate.  During this second stage, the neutral
fraction continues to increase, but it lags far behind the Saha prediction.  Although thermodynamic equilibrium no longer applies in 
this redshift range, the lifetime $\beta_{\mathrm{Li}}^{-1}$ of a neutral lithium atom is still small compared to the Hubble time 
$H^{-1}$.  Thus a ``steady state'' occurs in which the net \LiI\ production rate
$df_{\mathrm{Li}}/dz$ is much smaller than either the recombination or photoionization rates
in Eq.~(\ref{eqn:rateequation}).  Thus we have
\begin{equation}
f_{\mathrm{Li}} \approx \alpha_{\mathrm{Li}} n_e /( \beta_{\mathrm{Li}} + \alpha_{\mathrm{Li}} n_e) 
\approx \alpha_{\mathrm{Li}} n_e / \beta_{\mathrm{Li}}.
\label{eqn:steadystate}
\end{equation}
[This steady-state solution for lithium is analogous to the ``ionization equilibrium'' of \HI\ and H$^+$ in the Lyman-$\alpha$ forest.  
In both cases the recombination rate to form the neutral species (\LiI\ or \HI) is counterbalanced by photoionization by UV photons
from a non-thermal source, which consists of hydrogen recombination radiation in the case of lithium versus stars and quasars in the
Lyman-$\alpha$ forest.  Neither case represents a true thermodynamic equilibrium because the radiation field is non-thermal.]

\subsection{Residual steady-state stage ($20<z<400$)}

When hydrogen recombination freezes out at $z\sim 800$, the residual ionization fraction $x_e\sim$few$\times 10^{-4}$ changes
very little.  The Lyman-$\alpha$ photons produced at $z\sim 800$ are cosmologically redshifted and by $z\sim 400$ are no longer 
capable of ionizing \LiI.  Thus one might naively expect that after $z\sim 400$, lithium would fully recombine to \LiI.  As can be 
seen from Fig.~\ref{figs:lirec}, this is not the case: even after hydrogen recombination is ``frozen out,'' there are still a 
few residual recombinations occurring.  These act as a source of UV photons, keeping the neutral lithium fraction small.  

During the period of residual recombination, the steady-state 
solution for the neutral fraction can be simplified into a scaling law by making two main 
assumptions: matter temperature changes adiabatically ($T \propto a^{-2}$) after thermal decoupling at $z\sim 200$, 
and the relic electron fraction $x_e$ is nearly constant after freeze-out.  The low 
temperature limit of the recombination rate coefficient in Eq.~(\ref{eqn:liradiativerecombination}) 
for adiabatic matter gives $\alpha_{\mathrm{Li}} \propto T^{-1/2} \propto a$.  The photoionization rate is then
\begin{equation}
\beta_{\mathrm{Li}} \propto \alpha_H n^2 x_e^2 t \propto a^{-7/2}.
\label{eqn:ionizingflux}
\end{equation}
($a \propto t^{2/3}$ for matter domination, $H \propto t^{-1}$. The number density squared 
scales as $n^2 \propto a^{-6}$.)  Likewise for hydrogen, $\alpha_{H} \propto a$, so that the 
neutral lithium fraction is given by,
\begin{equation}
f_{\mathrm{Li}} = \frac{ n_{\mathrm{Li}} }{ n_{\mathrm{Li}^+} + n_{\mathrm{Li}} } \approx \frac{ 
n_{\mathrm{Li}} }{ n_{\mathrm{Li}^+} } = \frac{ \alpha_{\mathrm{Li}} }{ \beta_{\mathrm{Li}}} 
n_e \propto \frac{a }{ a^{7/2} } a^{-3} \propto a^{3/2}.
\label{eqn:neutralfraction}
\end{equation}
The $f_{\mathrm{Li}} \propto a^{3/2}$ scaling law is shown in Fig.~\ref{figs:lirec} and does a remarkably good job of reproducing the 
full numerical solution in the range from thermal decoupling at $z\sim 200$ to lithium freeze-out at $z\sim 20$.

\subsection{Freeze-out ($z<20$)}

As the universe continues to expand, the number of residual recombinations drops.  Consequently the UV background becomes 
fainter, and eventually the lifetime of a lithium atom exceeds the Hubble time, $\beta_{\mathrm{Li}}^{-1}\gtrsim H^{-1}$.  Here, the 
neutral fraction of lithium breaks away from the $a^{3/2}$ scaling and approaches a constant (see Fig.~\ref{figs:lirec}).

At some point the freeze-out is interrupted by cosmic reionization caused by the first stars and quasars.  Current observations
\cite{2002AJ....123.1247F} indicate that hydrogen was reionized at $z\ge 6$; lithium was presumably ionized even earlier because the
pre-reionization universe was transparent to photons with energies of 5.4--13.6 eV (except in the Lyman lines), which are capable of
ionizing lithium.  Regardless, this redshift range is not of interest for observations of the lithium resonance line scattering
signal: scattering at $z=20$ would be observed at a wavelength of 14~$\mu$m, where the CMB radiation is many orders of magnitude 
fainter than foreground emissions.

\section{Signature in the CMB \label{sec:signature}}

The observability of the lithium signature in the CMB is determined principally by the optical depth through the \LiI\ doublet.  As
noted in previous work \cite{2002ApJ...580...29S, 2002ApJ...564...52Z, 2001ApJ...555L...1L}, the major effect of lithium on small
angular scales is to suppress the CMB temperature and polarization anisotropies by a factor of
$\exp(-\tau_{\mathrm{Li}})$, where $\tau_{\mathrm{Li}}$ is the optical depth through the \LiI\ doublet.  This suppression factor
depends on the scattering redshift $z_s$, which in turn depends on the wavelength of the observation through the usual relation
$z_s=\lambda_{\rm obs}/(6708$\AA$)-1$.  There are also intermediate-scale effects ($\ell\sim 100$), namely the Doppler temperature
anisotropy due to motion of the gas at the lithium scattering redshift, and the polarization induced by \LiI\ scattering.  In the
optically thin regime $\tau_{\mathrm{Li}}\ll 1$ of interest here, the magnitude of the effect as measured by, e.g. the change $\Delta
C_\ell$ in the CMB power spectrum, is proportional to $\tau_{\mathrm{Li}}$.  This section is primarily concerned with computing 
the optical depth $\tau_{\mathrm{Li}}$.  We will then argue by scaling from the results of Stancil {\em et~al.} 
\cite{2002ApJ...580...29S} that the signature is too small to be observed.

The optical depth of neutral lithium in terms of the neutral fraction, neglecting the 
small fraction occupying the upper states is~\cite{2002ApJ...580...29S, 1997MNRAS.288..638B, 
2000ApJS..128..407S}
\begin{equation}
\tau_{\mathrm{Li}}(\lambda_{\rm obs}) = \sum_i \frac{\lambda_{1i}^3 A_{i1} }{8 \pi} \frac{g_i}{g_1}
\frac{ f_{\mathrm{Li}}(z_{s,i}) X_{\mathrm{Li}} n_H(z_{s,i}) }{ H(z_{s,i})}.
\label{eqn:liopticaldepth}
\end{equation}
Here the sum extends over all excited levels accessible from the ground level, and the scattering
redshifts are given by $z_{s,i}=\lambda_{\rm obs}/\lambda_{1i}-1$.  Among those, 
only the $2p$ $J=1/2$ and $J=3/2$ excitations at $1.8478$~eV contribute for photons at low 
frequencies (the next allowed excitation is to $3p~^2P_{1/2,3/2}$ at $3.7$~eV, where dust 
complicates observations), and each has an Einstein coefficient of $3.72 \times 
10^7$~$s^{-1}$, wavelengths of $\lambda_{1/2} = 6707.76$~\AA~and $\lambda_{3/2} = 
6707.91$~\AA, and degeneracies $g_{1/2} = 2$ and $g_{3/2} = 
4$~\cite{2002ApJ...580...29S,2003IAUJD..17E..13M}.  The degeneracy of the ground level is $g_1=2$.

\begin{figure}
\epsfxsize=3.3in
\begin{center}
\epsffile{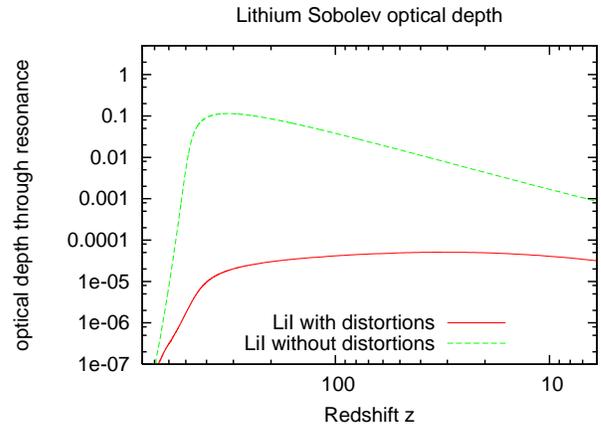}
\end{center}
\caption{A comparison of the optical depth through the resonance of neutral lithium, with and without the ionizing 
flux from hydrogen recombination.} \label{figs:opticaldepth} \end{figure}

We also need to know the lithium abundance $X_{\mathrm{Li}}$ in order to compute the
optical depth.  There is some discord about the value of $X_{\mathrm{Li}}$.  Recent standard 
BBN abundance calculations based on the R-matrix method for low energy nuclear cross sections yield 
$X_{\mathrm{Li}} = (4.15^{+0.49}_{-0.45}) \times 10^{-10}$~\cite{2005NuPhA.752..522C}, while 
observations of low-metallicity stars give $X_{\mathrm{Li}} = (1.23^{+0.68}_{-0.32}) \times 
10^{-10}$~\cite{2000ApJ...530L..57R}.  The BBN value of $X_{\mathrm{Li}} = 4.15 \times 
10^{-10}$ is used throughout this analysis.  (If the measurement based on stars is correct,
the optical depth is correspondingly reduced, $\tau_{\mathrm{Li}}\propto X_{\mathrm{Li}}$.)
The optical depth with and without the spectral distortions is shown in Fig.~\ref{figs:opticaldepth}.

The optical depths shown here are 3--4 orders of magnitude smaller than those found by Stancil {\em et~al.} 
\cite{2002ApJ...580...29S} in their Model A (which most closely matches the cosmology used here) in the interesting redshift range 
$200<z<500$.  Instead of a $\sim 20$\% change in the CMB power spectrum at $\lambda=268\,\mu$m ($z_s=400$), we have 
$\tau_{\mathrm{Li}}\sim 10^{-5}$ and find a fractional change of the order of $|\Delta C_\ell/C_\ell|\sim 2\times 10^{-5}$.  A 
similar conclusion holds at other redshifts because we find that $\tau_{\mathrm{Li}}$ is always $<10^{-4}$.  Thus, with the 
standard BBN abundance of lithium, the temperature or polarization due to the lithium feature never exceeds about $10^{-4}$ of the 
primary CMB anisotropy.  This is extremely difficult to measure even in principle, and is well beyond the capabilities of current 
experiments in the relevant wavelength range.  

In practice, the lithium signature must also be disentangled from the far infrared 
background (FIB).  Zaldarriaga \& Loeb \cite{2002ApJ...564...52Z} estimate that the FIB fluctuations are comparable to the 
lithium signal in polarization and dominate over lithium in temperature anisotropy at $z_s\sim 500$, if $\tau_{\mathrm{Li}}$ is of 
order unity.  At lower $z_s$, the FIB-to-CMB ratio increases dramatically.  Since we calculate $\tau_{\mathrm{Li}}<10^{-4}$, it 
follows that the lithium signal should be at most $\sim 10^{-4}$ of the FIB fluctuations.  Thus the lithium signal will be 
undetectable unless both the experimental capabilities improve dramatically, and the FIB fluctuations turn out to be much 
smaller than predicted or very simple to model.

It is conceivable that the lithium abundance is greater than that predicted by standard BBN, with some inhomogeneous models giving 
$X_{\mathrm{Li}}$ one to two orders of magnitude larger.  This is only sufficient to bring $\tau_{\mathrm{Li}}$ up to 
the $10^{-3}$--$10^{-2}$ range, which is still much smaller than previous predictions.

\section{Subtleties \label{sec:subtle}}

While the three-level approximation provides a convenient way to estimate the spectral distortion from hydrogen recombination, it is a
non-standard application of an approach that was developed largely to calculate the free electron fraction accurately.  In this
section, we justify its application to the problem of lithium subject to concerns about lithium feedback, i.e. the possibility that
\LiI\ photoionizations could absorb the spectral distortion (Sec.~\ref{ss:1}), \HI\ $2s$ and $2p$ falling out of equilibrium at late
times, influencing the production of the distortion (Sec.~\ref{ss:2}), and the accuracy of case B hydrogen recombination at late
times (Sec.~\ref{ss:3}).

\subsection{Lithium feedback}
\label{ss:1}

\begin{figure}
\epsfxsize=3.3in
\begin{center}
\epsffile{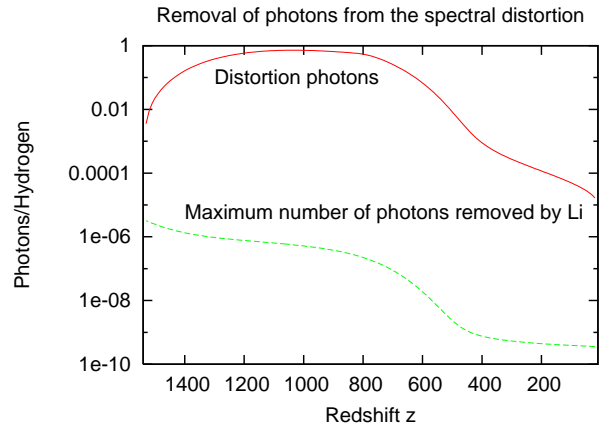}
\end{center}
\caption{The maximum number of photons that lithium recombinations can remove from the
distortion population is significantly less than the total distortion population, based on $X_{Li} = 
4.15 \times 10^{-10}$.  The distortion spectrum is well-approximated by just those photons which are 
generated by hydrogen, without lithium feedback.}
\label{figs:eating} \end{figure}

As lithium recombines to the ground state, it absorbs photons from the distortion field.  In
the most extreme case of this, all of the distortion photons are absorbed by a small number
of lithiums, accelerating lithium recombination.  We would like to show that this ``lithium feedback'' is 
negligible, and that it is realistic to use the distortion field set up by hydrogen recombination alone. 
This can be shown by checking that the total number of photons absorbed by lithium that could have ionized a \LiI\ atom at redshift 
$z$ is much less the total number of distortion photons.  The number of photons per H nucleus that are absorbed 
($X_{\mathrm{removed}}$) is equal to, and bounded by, 
\begin{eqnarray}
X_{\mathrm{removed}} &=& \int_{z}^{z_e(z)}\frac{ X_{\mathrm{Li}} (1-f_{\mathrm{Li}}) n_e \alpha_{\mathrm{Li}} }{ H(z) }
\frac{ dz }{1+z} \nonumber \\ &+& X_{\mathrm{Li}} (f_{\mathrm{Li}}|_{\mathrm{initial}}-f_{\mathrm{Li}}|_{\mathrm{final}})
\nonumber \\
&\le&  X_{\mathrm{Li}} \left [ \int_{z}^{z_e(z)}\frac{ n_e \alpha_{\mathrm{Li}} }{ H(z) } \frac{ dz }{1+z} + 1 \right ] ,
\label{eqn:eatinglithium}
\end{eqnarray}
where the endpoint of the integral is the maximum redshift where photon from a Lyman-$\alpha$ decay 
can be created so that it reaches $z$ with enough energy to ionize lithium,
\begin{equation}
z_e(z) = \frac{ E_{Ly\alpha} }{ \chi_{\mathrm{Li}} } (1+z) -1.
\label{eqn:endpoint}
\end{equation}
By comparison, the number of photons above the \LiI\ ionization threshold in the distortion per H nucleus ($X_{\mathrm{field}}$) is
\begin{equation}
X_{\mathrm{field}} = \int_{\chi_{\mathrm{Li}}}^{E_{Ly\alpha}}\frac{d E' }{ E'} r_{distortion}(E',z).
\label{eqn:numberindistortion}
\end{equation}
The two integrals, Eqs.~(\ref{eqn:eatinglithium}) and (\ref{eqn:numberindistortion}), are shown in
Fig.~\ref{figs:eating}.  The number of photons removed from the distortion is
several orders of magnitude less than the total distortion population.  This confirms that absorption of
the distortion photons by lithium can be neglected.

\subsection{2s--2p non-equilibrium}
\label{ss:2}

\begin{figure}
\epsfxsize=3.3in
\begin{center}
\epsffile{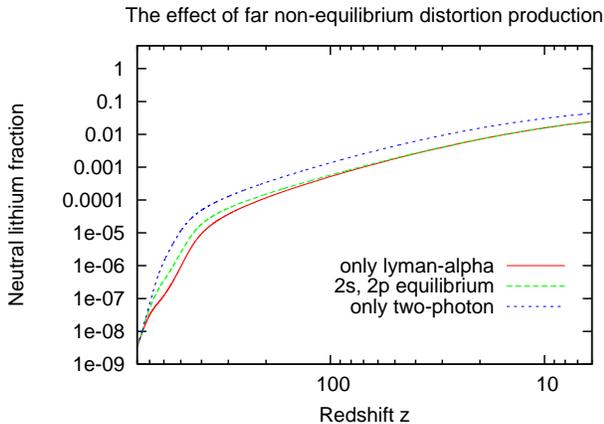}
\end{center}
\caption{For a spectral distortion produced by only Lyman-$\alpha$ decays, the spectral distortion is 
much harder, and lithium recombines less readily at early times.  Alternately, for a 
distortion produced by only two photon decays, the softer radiation allows lithium to 
recombine more efficiently.  This simulates the effect of hydrogen $2p$ and $2s$ falling out 
of equilibrium in its most extreme form.  In both cases, lithium remains largely ionized up to 
late times.}
\label{figs:noneq} \end{figure}

The second primary concern is that the spectral distortion derived from rates in the 
three-level approximation may not be accurate because it assumes that $2s$ and $2p$ are in 
equilibrium in hydrogen.  The reactions that keep $2s$ and $2p$ in equilibrium involve absorption and emission of Balmer and Paschen 
line photons.  These are expected to slow down not long after hydrogen 
recombination when the CMB photons redshift to sufficiently low energies that they cannot excite Balmer transitions.  This means that 
$2s$ and $2p$ could fall out of equilibrium because the ratio of effective recombination
rates $\alpha^{eff}_{2p}/\alpha^{eff}_{2s}$ differs from the statistical ratio $g_{2p}/g_{2s}=3$.
This becomes a concern for how well we understand the spectral distortion.  If at late times, the 
majority of the decays are via the two-photon mechanism, then the ionizing flux will not be as hard as when 
production is a mixture of two-photon and Lyman-$\alpha$, or all Lyman-$\alpha$.  Here we look at the 
worst and best case scenarios where the distortion is produced by only two photon processes, 
or only Lyman-$\alpha$ processes, shown in Fig.~\ref{figs:noneq}.  This is a litmus test for 
whether $2s$ and $2p$ being out of equilibrium has any bearing on the conclusion: the lithium 
ionization history lies somewhere between both of these cases.  As can be seen from the figure,
even the two extremes do not alter the conclusion that lithium recombination is dramatically inhibited.

\subsection{Accuracy of recombination coefficients}
\label{ss:3}

The third possible concern is that the case B hydrogen recombination coefficients 
used in {\verb recfast } from Pequignot et al.~\cite{1991A&A...251..680P} will not scale properly below $40$~K, where 
residual hydrogen recombinations can ionize lithium (the coefficients are accurate at high temperatures during the bulk of 
recombination)~\cite{1994MNRAS.268..109H, 1996ApJS..103..467V}.  The case A hydrogen recombination rates in Verner and 
Ferland~\cite{1996ApJS..103..467V} are accurate down to $3$~K, with a maximum error of $6\%$.  These can be converted to accurate 
case B rates that are similarly accurate down to $3$~K by subtracting recombinations to the $1s$ level.  The fitting form for 
case B hydrogen recombination of Pequignot et al. used in {\verb recfast } departs from this more accurate calculation by less 
than $\sim 10\%$ for temperatures above $3$~K.  The recombination coefficients used in {\verb recfast } are accurate to several percent 
over the redshift range in question.

\section{Conclusions \label{sec:conclusions}}

We have shown that the non-thermal UV flux from hydrogen recombination is sufficient to keep lithium almost completely ionized
throughout the cosmic Dark Ages. Previous studies have only considered the thermal tail of the CMB, which is not enough to ionize
lithium at times later than $z \approx 400$--$500$.  The UV flux from hydrogen recombinations dramatically reduces the optical depth
through the \LiI\ doublet and the prospect for observing a lithium signal in the CMB.  This is unfortunate because the lithium
scattering signal, had it been detectable, would have been a valuable cosmological probe in the $200<z<500$ redshift range and would
have improved our understanding of BBN physics by permitting a clean measurement of the primordial lithium abundance.

UV radiation from hydrogen recombination may also be important for the production of molecules such as H$_2$,
LiH, and HD because some of the the intermediate products, catalysts and final states can be destroyed by UV radiation.  H$_2$ 
and HD have been suggested as coolants for gas clouds in the early universe which could be important for the formation of the 
first baryonic structures.  Cooling processes will also be crucial in our understanding of upcoming high-redshift 21~cm surveys, 
both directly if an \HI\ signal from minihaloes is observed \cite{2005ApJ...624..491I}, and indirectly because cooling affects 
the formation of UV sources such as stars.  These results highlight the importance of developing a code to track the radiation 
field during recombination, and we hope to discuss a new algorithm for this in a future paper.

\acknowledgments

C.H. acknowledges the support of NASA grant NGT5-50383.  We acknowledge useful discussions with Uro\v s Seljak.

\bibliography{cosmo}

\end{document}